\documentclass{article}

\usepackage{arxiv}

\usepackage[utf8]{inputenc} 
\usepackage[T1]{fontenc}    
\usepackage{hyperref}       
\usepackage{url}            
\usepackage{booktabs}       
\usepackage{amsfonts}       
\usepackage{nicefrac}       
\usepackage{microtype}      
\usepackage{lipsum}
\usepackage{mathtools}
\usepackage[norelsize,ruled,linesnumbered]{algorithm2e}
\usepackage[
backend=biber,
style=ieee,
sorting=ynt
]{biblatex}
\usepackage{url}
\usepackage{microtype}

\usepackage{fancyhdr}
\usepackage{csquotes}

\title{On the Fairness of \emph{Fake} Data in Legal AI}

\author{
 Lauren Boswell\thanks{Thanks to Angus McCrabb for editing} \\
  School of Law\\
  University of Sydney\\
  \texttt{lbos5805@uni.sydney.edu} \\
   \And
  Arjun Prakash \\
  School of Computer Science\\
  University of Sydney\\
  \texttt{apra3637@uni.sydney.edu.au} \\
  \AND
}

\begin{document}
\maketitle
\begin{abstract}
The economics of smaller budgets and larger case numbers necessitates the use of AI in legal proceedings.  We examine the concept of disparate impact and how biases in the training data lead to the search for fairer AI. This paper seeks to begin the discourse on what such an implementation would actually look like with a criticism of pre-processing methods in a legal context . We outline how pre-processing is used to correct biased data and then examine the legal implications of effectively changing cases in order to achieve a fairer outcome including the black box problem and the slow encroachment on legal precedent. Finally we present recommendations on how to avoid the pitfalls of pre-processed data with methods that either modify the classifier or correct the output in the final step. 
\end{abstract}


\section{Introduction}
Artificial intelligence utilized in the legal system, and governance generally, is no novel concept in 2020. The economics of smaller budgets and larger case numbers means that it is more and more tempting for governments to look to automated solutions for decision-making processes \cite{Zavrnik2020}. Some states in the United States have deployed AI to calculate risk-assessments in the criminal justice system \cite{epic}. Australia has its sights on moving towards almost complete automatic processing of low-risk visa applications \cite{visa}. And where AI is not implemented, government agencies are developing umbrella ethical infrastructures for its imminent arrival \cite{NZAI,EUAI,CSIRO,UNESCO}. 

But championing efficiency and cost-reduction through the implementation of AI is not without its shortfalls. The well-documented case of racial bias in risk-assessment algorithms used to guide judges when determining bail and sentencing in the United States sounded the alarm that if AI is to take on a quasi-judicial role in the legal system it must be responsible, and it must be \emph{fair} \cite{doyle_barabas_dinakar_2019,propublica}. 

It seems that there is an institutional demand that where AI is used in government, it must be transparent, explainable, and fair \cite{NZAI,EUAI,CSIRO, UNESCO}. But the cross-over of AI from a proprietary use to governmental implementation is still nascent, so the ethical discussion is just that––a discussion. Arguably, this paper is also guilty. Conversely, in the machine learning community the discourse revolves around satisfying various statistical definitions of fairness, many of which are competing or are mutually exclusive. The practical considerations of AI implementation at a societal level is generally relegated to an afterthought behind the empirical results. The axioms of statistical fairness and the axioms of legal fairness seemingly exist in two separate vacuums unaware of each other. 

This paper wants to begin the discussion of what the practical implementation of fair AI in the legal system looks like. In particular, we examine a popular class of methods that achieve fairness by modifying the original data. These methods have appeared in the most elite machine learning journals in recent years and continue to propagate the discourse on fair AI \cite{Calmon2017,Kamiran2009,johndrow2019}. Our intention is to investigate these methods and examine if they are even compatible with principles such as the rule of law. First, we will introduce the concept of \emph{disparate impact} and explain why pre-processing methods are needed. Then we explain two of the most influential methods, \emph{data massaging} and \emph{otimised pre-processing}, before briefly highlighting some of the more recent developments using \emph{SMOTE} and \emph{Optimal Transport Theory}. Finally, we aim to balance these pre-processing methods against common law values in a general context. 

\section{Disparate Impact}
The motivation for fair machine learning is because of \emph{disparate treatment} and \emph{disparate impact}. \emph{Disparate treatment} refers to the explicit discrimination of a protected class. Far more insidious is \emph{disparate impact}, which occurs when a protected class is discriminated against indirectly \cite[1]{Feldman2015}. When decisions are made by an algorithm, especially a black box, these kinds of biases can be difficult to uncover. Unfortunately, the data mining process itself is fraught with a number of traps that inevitably lead to the disparate impact. Issues in data collection, data labelling and proxy variables \cite[677-693]{Barocas2016} can all cause members of a protected class to suffer unintentional discrimination.

For example, predictive policing can be a source of biased training data, as it satisfies the notion of a self-fulfilling prophecy \cite[5]{dunkelaufairness}. As certain areas are targeted for more policing, more crime is found in these areas. As a result, features associated with these areas become more prevalent in the data, which in turn informs the algorithm of where crime is to be found. This cycle amplifies prejudicial practices and can even serve as historical justification for future discrimination. 

\emph{Proxy variables} are particularly difficult as they introduce systematic bias even when the protected features in question are removed. For example, an algorithm that is not permitted to use race as a variable in its decision but is permitted to use a zip code still introduces disparate impact, because while it removes obvious racial bias, it fails to appreciate how a zip code stratifies socio-demographic groups and therefore implicitly reintroduces the protected race variable back into the data  \cites{Prince2020}. More generally, proxies occur when the criteria of making a genuine decision also happens to be indicative of class membership \cites[691]{Barocas2016}. 

Continuing with the predictive policing example, it also highlights the issue of proxy variables. For example, the Los Angeles Police Department uses a point system to identify the worst offenders \cite{selbst2017disparate}[138]. The LAPD adds a point to a person's profile per police contact, leading to the very same feedback loop, particularly in minority neighbourhoods. While neighbourhood is not directly a legally protected variable, it serves as a proxy for racial or socio-demographics \cite[5]{dunkelaufairness}, meaning that the any algorithm can implicitly learn that a person of color is more likely to be a criminal \cite[138]{selbst2017disparate}.

These issues, particularly of proxy variables, motivate the need for more sophisticated methods to handle disparate impact. The idea of pre-processing is to actually remove the underlying biases from the training data. This \emph{repaired} training data is then used to train the machine learning model resulting in a fair classifier.

\section{Pre-processing}
In this context, we refer to \emph{pre-processing} as the modification of the training data before it is used to train some predictive model. The motivation for such a step is simple; given the initial data set may already be biased, why not \emph{repair} this data by removing this bias and then train the model on this modified repaired data set. Below we outline two of the most cited pre-processing methods published in \emph{NIPS} \cite{Calmon2017} and \emph{ICCCM}\cite{Kamiran2009} to combat these implicit biases. Although a fuller review of literature can be found in the surveys of \citeauthor{friedler2019} and \citeauthor{dunkelaufairness}, we will attempt to make this technical jargon more accessible to a broader audience by introducing fictitious, rudimentary examples to explain the algorithmic yields in the spirit of algorithmic transparency. These examples seek to highlight the tension between \emph{group fairness}; the probability for an individual to be assigned the favourable outcome to be equal across the privileged and unprivileged group, \emph{individual fairness}; similar groups need to be treated similarly and compatibility with the common law.

\subsection{Massaging}
\label{sec:massage}
The early work of \cite{Kamiran2009} seeks to massage the training data to create an unbiased data set by flipping the labels of certain data points in order to satisfy group fairness. The authors find that group fairness is improved with minimal impact on accuracy.

\begin{algorithm}[H]
\SetAlgoLined
\SetKwInOut{Input}{Input}
\SetKwInOut{Output}{Output}
\Input{Data, Class Label, Protected Attribute}

Rank each sample relative to how likely it is to be in the positive class;

Create a descending ordered list of the samples that were in the protected class and true label was negative, \emph{pr};

Create an ascending ordered list of the samples that were not in protected class and true label was positive, \emph{dem};

Calculate M, the minimum number of modifications needed

\While{$count < M$}{
    Select the top object in \emph{pr} and flip the label;
    
    Select the top object in \emph{dem} and flip the label;
    
    remove the top elements from both \emph{pr} and \emph{dem};
    increment $count$
}

\Output{Massaged data}
\caption{High Level Algorithm of \citeauthor{Kamiran2009}}
\end{algorithm}

This algorithm in a legal context is actually relabelling the data. To illustrate massaging, we introduce our example of a new case to be classified, \emph{Plato}, who is a not a member of a protected class. 

In trying to satisfy the principle of individual fairness, paradoxically, individual fairness is actually violated. Aristotelian Equality tells us that like cases should be treated alike. In the original data, \emph{Plato} could have been compared to a set of cases, \emph{S}, that were classified with the positive outcome. However, after the data is massaged, the same set of cases, \emph{S}, have their labels switched to the negative outcome. As a result, \emph{Plato} suffers the negative classification.

\subsection{Optimised Pre-processing}
\label{sec:opp}
More recently \citeauthor{Calmon2017} formulated an  optimisation to transform the dataset $\mathcal{D} = \{X_i,Y_i,Z_i\}$ into $\{\tilde{X_i},\tilde{Y_i}\}$. Where $X$ is the feature vector, $Y$ is the label vector and $Z$ is the vector of protected features using a randomised mapping of $p_{\tilde{X}, \tilde{Y} \mid X, Y, Z}$

This is done by solving for the following problem:

$\begin{aligned} \min _{p_{\tilde{X}, \tilde{Y} \mid X, Y, Z}} & \Delta\left(p_{\tilde{X}, \tilde{Y}}, p_{X, Y}\right) \\ \text { s.t. } & D\left(p_{\tilde{Y} \mid Z}(y \mid z), p_{Y_{T}}(y)\right) \leq \epsilon_{y, z} \text { and } \\ & E(\delta((x, y),(\tilde{X}, \tilde{Y})) \mid Z=z, X=x, Y=y) \leq c_{z, x, y} \forall(x, y, z) \in \mathcal{D} 
\end{aligned}$

$p_{\tilde{X}, \tilde{Y} \mid X, Y, Z}$ is a valid distribution
\newline

Here $\Delta(.,.)$ is a dissimilarity measure between distributions to be minimised. $D(.,.)$ is an arbitrary distance measure \footnote{enforcing the constraint that the distributions over outcomes, conditioned on protected attributes should be as similar to each other as possible}, $\delta(.,.)$ is a measure of distortion \footnote{enforcing the constraint that any individuals information should not change much}. This kind of formulation is unique as it addresses the problem of disparate impact as an optimisation problem and acknowledges the trade off between different kinds of fairness. 

The key issue here is that $c_{z, x, y}$ is a parameter for controlling distortion. This means for any value of $c_{z, x, y} > 0$, it is permissible to have small perturbations in this new transformed data set, which is then used to train the model. It is also unclear how to tune $c_{z, x, y}$. In practice, it seems like an implementor of this method could arbitrarily set the value without any legal oversight.

In optimized pre-processing, \emph{Plato}, is a member of a protected class. Suppose \emph{Plato's} case turns on a particular set of facts, such as a number of previous convictions. Again, in trying to satisfy individual fairness, individual fairness is violated. In the original data $\{X_i,Y_i,Z_i\}$, \emph{Plato} could have been compared to a set of cases, $S_1$, that had similar counts of previous convictions. Recall $X$ is the vector of facts of the cases, $Y$ is the vector of outcomes of the cases, and $Z$ is the vector the protected features. The pre-processing is applied to remove the presence of $Z$ from the data to combat disparate impact. This creates a new randomised dataset $\{\tilde{X_i},\tilde{Y_i}\}$.  After the data is pre-processed, \emph{Plato} would now be compared to a new set of cases $S_2$ which have been randomly generated \footnote{Albeit with statistical justification} to satisfy the above optimisation. When $S_2$ is used to train the algorithm, it means that these modified facts, like the number of previous convictions, were used to determine the outcome of \emph{Plato's} case. These facts never existed outside of the algorithm. Furthermore, the choice of the value of  $c_{z, x, y}$ has the ability to greatly affect the outcomes of cases if not selected carefully, and the fact that no guidance is offered by the authors on how to choose the value of $c_{z, x, y}$ is concerning.

\subsection{Other Methods}
Pre-processing biased data continues to be a popular area in the machine learning community, particularly using \emph{optimal transport theory}. \citeauthor{pmlr-v97-gordaliza19a} propose a method to remap the training data $X$ to a modified $\tilde{X}$ such that the conditional distribution with respect to the protected attribute $Z$ is the same for different values of $Z$. In similar work, \citeauthor{johndrow2019} also use transport theory to transform the original dataset such that all information about a potential protected variable is removed by actually achieving pairwise independence between each $X_i$ and protected variable $Z$. More recently, \citeauthor{Wang2019} aims to improve a classifier by learning the counterfactual distributions in the data, and then perturb the data to minimise disparate impact on non-baseline groups. "For example, if $X_i$ is the number of prior arrests and $Z$ is race, a pairwise adjustment would result in a race-adjusted measure of the number of prior arrests" \cite[8]{Wang2019} . For \emph{Plato}, this means a similar problem as section \ref{sec:opp}. 

In other recent work, simply generating pseudo-instances of minority classes using SMOTE has also been shown to give promising results \cite{iosifidis2018dealing}. For \emph{Plato}, this would mean potentially being compared to cases that are completely randomly generated, both the facts and outcomes - exacerbating the problems from both sections \ref{sec:massage} and \ref{sec:opp}.

Further summaries of pre-processing methods can be found in the works of \citeauthor{friedler2019} and \citeauthor{dunkelaufairness}.

\section{Legal \& Social Considerations}

In preparing an argument against pre-processing, it was difficult to build an argument with relevant jurisprudence, because the idea of repairing data, i.e. changing the facts or decisions of cases extra-judicially, is so absurd that it is unlikely to be a concept in need of a theoretical framework. Abdul Paliwala, of the University of Warwick, is critical of the inadequate jurisprudence in the age of AI, stating that 

\begin{displayquote}
"Without a proper awareness of key issues, it is possible that these systems will either replicate past failures or result in systems which though successful in a technical sense produce results which not advance the need for proper legal development"\cite{paliwala_2016}.  
\end{displayquote}

The authors of this paper tend to agree, but will advance a very rudimentary legal critique of pre-processing methods, which as the jurisprudence in this area progresses, might later become a more focussed discussion on why pre-processing is counter-intuitive to the current notions of legal ethics.

One of the most obvious criticisms of the pre-processing methods discussed above is if there is even any legality to repairing data used in AIs by machine learning practitioners. In common law jurisdictions, it is antithetical to so much as contemplate the idea that once a matter is decided, it is liable to be changed without proceeding through the proper mechanisms of appeal. So much so, that it goes against a Dicey conception of the rule of law, which is a set of principles on how governments should govern\cite{jowell_oliver}. If previous decisions are potentially altered by an AI to achieve empirical fairness, people accessing the judicial system might struggle to understand the case they must meet if previous decisions are altered at random in an endless pursuit of statistically-perceived fairness. There is no certainty of the law\cite{jowell_oliver}, because the repaired data could hypothetically be the reason a case is decided in the negative if AIs are heavily relied on within the judicial decision, be it a sole decision-maker or as supplementary tool for judges.

While AIs have the ability to satisfy the 'efficiency' aim of the rule of law, it appears to come at the cost of accountability. Where machine learning practitioners possess the ability to perturb legal data to accomplish a perceived measure of fairness, they do so in a language not understood by the body politic. Where the machine learning practitioners, for all intents and purposes, quasi-make the rules by repairing data, they should be subjected to the scrutiny of those whose fidelity is paramount\cites[11]{}{jowell_oliver}. It does not follow that these solutions should be implemented obscured or without inquiry, or they will directly contravene the rule of law. 

Adrian Zuckerman, Emeritus professor of civil procedure at the University of Oxford purports that, “AI decision-making may lead to an ever-widening gulf between machine law and human conceptions of justice and morality, to the point where legal institutions cease to command loyalty and legitimacy”\cite{zuckermann_2020}. The authors of this paper agree. If pre-processing methods are utilized in AIs making decisions, they have the power to make legal change. By this we do not imply that there will be instantaneous chaotic reversals of long-venerated precedent. Rather, the repaired data on which the AI has trained on will become part of the infamous 'black box,' and it will be nearly impossible to know when a decision-making AI is presented with a matter for decision whether or not the turning point of the AI’s outcome was based on repaired data or the original data. We posit that the repaired data will create incremental, perhaps imperceptible changes to precedent at first.  And as AIs advance and humans become more trustful of them, the judicial system might be likely to defer to an AI's judgment, acquiescing either knowingly or unknowingly to a shift from the common law’s axiom of 'fairness as morality' to an empirical axiom. 

Finally, we have concerns over how such pre-processing methods might be received in the public discourse. \emph{Fake news} already competes with reality in the current zeitgeist. Cognitive dissonance paves the way for charged emotional appeals and attacks on feelings and beliefs \cite{kaplan_2019}. In an era where new AI innovations such as deep fakes and language models are capable of exacerbating the problem \cite{Caldwell2020}, we believe that pre-possessing methods may get caught up in the hysteria. The very notion of deciding fates or affecting rights based off of \emph{fake data} is a novel 'Black Mirror-esque' idea and such stories are primed to be sensationalised causing even more distrust in the legal machinery.

\section{Recommendations}
The old adage, "just because you can, doesn't mean you should" might be best followed in this context, but in the age of AI, it is unrealistic to preclude AI from advancing into the legal system. We do not think fair AI is not possible in a legal context––we are just hesitant to recommend pre-processing methods as the means to a fair end. Methods such as \emph{in-processing} and \emph{post-processing} offer useful, more transparent alternatives to achieving fairer AI. 

In-processing refers to the technique of modifying the algorithm so that it can handle a biased dataset. In particular, we endorse methods that leave the training data unchanged and instead opt to use regularisation. In this way, it becomes explicit that the algorithm is actually penalised for taking protected features into account. For example, \citeauthor{kaishima2012} modify a logistic regression and add a penalty to the loss function in order to reduce discrimination learned by the model which "enforces a classifier’s independence from sensitive information" \cite[15]{kaishima2012}. Another interesting implementation of in-processing is an \emph{adversarial neural network} \cite{beutel2017data}. Here two neural networks are used with a share hidden layer. One, \emph{A}, learns to predict the outcome, the other, \emph{B}, learns to predict the value of the protected variable. By maximising the objective function of \emph{A} and minimising the objective function of \emph{B}, the algorithm is \emph{discouraged} from using the protected variables. Furthermore, the formulation of the objective function makes explicit the trade-off between accuracy and fairness. 

Post-processing methods take a possible biased classifier, trained on a possibly biased dataset and correct the output. \citeauthor{kamiran2012decision} designs a method that operates on cases that are close to the decision boundary of an algorithm. Simply, if a case is close to the boundary, and from the unprivileged class, the case receives the positive outcome (vice versa for the prevailed class). In the same paper, \citeauthor{kaishima2012} also proposes an ensemble based approach where a number of classifiers are used to make a decision. If all classifiers unanimously agree, then the decision is final. If at least one classifier disagrees, the unprivileged case receives the positive outcome while the prevailed class would receive the unfavorable outcome. 

In our opinion, both in-processing and post-processing offer reasonable alternatives to pre-processing. This is because the aforementioned methods do not modify the data and hence do not contribute to the incremental, perhaps imperceptible changes to precedent.  In the case of in-processing, the regularisation term is a more explicit formulation for enforcing the classifier's avoidance of protected features. Post-processing basically offers a \emph{safety net} that is able to catch problematic cases and, given careful consideration, is able to make the transparent corrections. 

The difference between pre-processing, in-processing and post-processing methods is subtle, and often times methods are amalgamated together into a \emph{fairness soup}, adding to the black box problem. More legal and technical scrutiny is needed, which is beyond the scope of our discussion. However, we believe that careful choices of in-processing and post-processing are worthy areas of investigation in the pursuit of fairer AI. 

Given that the economics dictate that AI play a larger role in judicial proceedings, we also have a set of general recommendations. We believe that AI should only be used in instances of positive rights, such as making decisions on the granting of licenses or travel visas. And if an AI is employed to make decisions that might affect a party to a claim, it should be constrained to small civil claims where the detriment to a party is nominal.

But if AI is implemented to make decisions judicially or extra-judicially, there should be express legislation to implement these AIs and their functions. We strongly believe that this issue should be accessible to the body politic, which would require thorough public consultation and positive engagement between legislators and voters to raise awareness around the issues of AI utilised in the legal system.

In this vein, we propose a possible system in which a decision-maker can be augmented by the AI. First, someone's matter would come before the court or decision-making body. The person must consent to the use of an AI for decision-making; if they do not, the decision will be made by a human. 

If the person consents to the AI and if the matter receives a positive outcome, by whatever algorithm is implemented, the outcome stands. If the case receives the negative outcome, it is re-evaluated by a human. The key here is that the case re-evaluator must not know that the matter has received the negative outcome from the algorithm to avoid any implicit bias that might be imputed from knowledge of the AI's decision. This can be achieved by simply adding the matter back into the pool of the people who did not consent. By carefully tuning the number of cases 'allowed' to be decided by the AI \footnote{which would be some fraction of the total number of matters}, it would be possible not to over-load the human re-evaluators with negative cases from the algorithm and a blind re-evaluation is possible.

\section{Concluding Remarks}
It is unlikely that AI will replace highest court justices. AI may never be able to have the 'intelligence' to piece together the ontologies of the legal system, or understand the contentious nature of individualised justice and Benthamite utilitarian public welfare considerations, to sufficiently serve its needs. But it is obtuse to assume that machine learning practitioners will not be up for the challenge. And where governments are looking towards leaner budgets and technology proprietors are tasked with that tall order, lawyers and policy makers must be able to keep pace. It is impractical to envisage a new generation of legal tech gurus and data science lawyers, but the two must not remain uninformed of each other if the two purport to influence the future of legal AI. It is imperative that the legal profession mobilise to figure out the questions to ask when it comes to legal AI, and we hope with this paper, we make these issues more accessible to both the legal and machine learning communities.

\printbibliography
\end{document}